\RequirePackage{ifpdf}
\ifpdf 
\documentclass[pdftex]{sigma}
\else
\documentclass{sigma}
\fi

\newcommand{\Tr}{\mathop{\mathrm{Tr}}\nolimits} 

\usepackage{hhline}  

\begin{document}
\allowdisplaybreaks

\renewcommand{\PaperNumber}{008}

\FirstPageHeading

\ShortArticleName{Status Report on the Instanton Counting}

\ArticleName{Status Report on the Instanton Counting}

\Author{Sergey SHADCHIN}
\AuthorNameForHeading{S. Shadchin}

\Address{INFN, Sezione di Padova {\rm \&} Dipartimento di Fisica 
``G. Galilei'', \\
Universit\`a degli Studi di Padova, via F. Marzolo 8, Padova, 35131, Italy}
\Email{\href{mailto:serezha@pd.infn.it}{serezha@pd.infn.it}} 

\ArticleDates{Received December 07, 2005, in final form January 18,
2006; Published online January 22, 2006}

\Abstract{The non-perturbative behavior of the 
${\mathcal N}=2$ supersymmetric Yang--Mills theo\-ries is both 
highly non-trivial and tractable. In the last three years 
the valuable progress was achieved in the instanton counting, 
the direct evaluation of the low-energy effective Wilsonian action of the theory. 
The localization technique together with the Lorentz deformation of the action provides 
an elegant way to reduce functional integrals, representing the effective action, 
to some finite dimensional contour integrals. These integrals, in their turn, 
can be converted into some difference equations which define the Seiberg--Witten 
curves, the main ingredient of another approach to the non-perturbative 
computations in the ${\mathcal N}=2$ super Yang--Mills theories. Almost all models 
with classical gauge groups, allowed by the asymptotic freedom condition 
can be treated in such a way. In my talk I explain the localization approach 
to the problem, its relation to the Seiberg--Witten 
approach and finally I~give a review of some interesting results.}

\Classification{81T60; 81T13}

\section{Introduction}

The goal of this presentation is to describe some aspects of the new 
approach to the non-perturbative computations in the supersymmetric 
extension of the Yang--Mills theory (${\mathcal N}=2$ super Yang--Mills, for the sake of brevity).

This model (with or without supplementary matter multiplets) 
possess a number of very interesting properties. First of all let us note 
that non supersymmetric Yang--Mills theory with the gauge group ${\rm U}(1)\times {\rm SU}(2) 
\times {\rm SU}(3)$ is the basis of the Standard Model, the model which describes real 
collider physics at the energies $\leq 100$~GeV (it is summer of 2005). Even though the perturbative 
computations fit well the experiment data, the non-perturbative 
information can still hardly be extracted from the theory. However, 
to describe correctly such effects as confinement we \emph{should} 
be able to go beyond the perturbative expansion. 

The problem given ``as is'' is too hard to be solved by existing methods. 
Instead, we can consider \emph{toy models} which are simpler to be solvable, 
and at the same time enable us to gain some intuition about the state of affair.
 One way to get such a toy model is to consider the supersymmetric extension of 
 the Yang--Mills theory. If we require the CPT-invariance in 4~dimensions 
 the following supersymmetric models are possible:
\begin{itemize}
\itemsep=0pt
\item \underline{${\mathcal N}=4$}: It is the supersymmetric 
extension of the Yang--Mills theory with maximal (renormalizable) supersymmetry. 
It has 8 supercharges. The perturbative effects come from the 1-loop diagrams, 
the non-perturbative effects are trivial. It is conformal theory, 
that means that the $\beta$-function is zero and the running coupling constant does not run.

\item \underline{${\mathcal N}=2$}: This supersymmetric 
extension of the Yang--Mills theory contains 4 supercharges. 
The perturbative effects are still 1-loop, but there are non-perturbative 
effects due to contribution of the instanton vacua. The non-perturbative 
effects such as confinement and the monopole condensation are present. 
At the same time the model contains the topological sector which allows 
us to compute explicitly all the non-perturbative contributions and therefore solve the model. 

\item \underline{${\mathcal N}=1$}: It is the simplest supersymmetric 
extension of the Yang--Mills theory. It might happen that it \emph{is} 
the realistic model. Experiments beyond 100~GeV will display if the 
Standard Model is just the supersymmetry broken version of the Minimally 
Supersymmetric Standard Model. This model is hard to solve, but some progress is achieved by 
the Dijgraaf--Vafa approach.

\item \underline{${\mathcal N}=0$}: At present the only realistic model, but not solved yet. 
\end{itemize}

From this list we see that the ${\mathcal N}=2$ super Yang--Mills theory is the only theory
 which is placed at the overlap of our possibility and our ambitions. 

Another motivation to study this theory is its connection to the \emph{String Theory}. 
More precisely to a theory which is believed to get rise of all superstringy models, 
so-called M-theory. At present the string theory computations are too hard 
to be performed by brute-force. However, some predictions of the M-theory are 
concerned to the ${\mathcal N}=2$ super Yang--Mills. Therefore if we
 have a method to compute the same quantities within ${\mathcal N}=2$ super Yang--Mills 
 itself, we can check indirectly the M-theory arguments and techniques.

This report was presented on 
the Sixth International Conference
``Symmetry in Nonlinear Mathematical Physics'' (June 20--26, 2005, Kyiv).

\section[Effective action and Seiberg-Witten solution]{Effective action and Seiberg--Witten solution}

In this section we describe the model, its low-energy effective action and the Seiberg--Witten solution for this model.

\subsection{Microscopic description}

The microscopic action is given by the following expression 
(we do not include matter multiplets for the sake of brevity):
\begin{gather}
S_{\rm micro}(X) = \frac{\Theta}{32\pi^2 h^\vee} \int {\rm d}^4 x \Tr F_{\mu\nu} \star F^{\mu\nu} \nonumber\\
\phantom{S_{\rm micro}(X) =}{}+ \frac{1}{g^2 h^\vee} \int {\rm d}^4 x \Tr 
\left\{ -\frac{1}{4} F_{\mu\nu} F^{\mu\nu} + 
\nabla_\mu H^\dag \nabla^\mu H - \frac{1}{2} {[H,H^\dag]}^2 \right. \nonumber\\
\phantom{S_{\rm micro}(X) =}{}
+ \left. i \psi^A \sigma^\mu \nabla_\mu \bar{\psi}_A - \frac{i}{\sqrt{2}} 
\psi_A [H^\dag, \psi^A] + \frac{i}{\sqrt{2}} \bar{\psi}^A [H,\bar{\psi}_A] \right\},\label{MicroAction}
\end{gather}
where $X = (A_\mu, H, \psi_A)$ is the set of fields, $A_\mu(x)$ being 
the gauge field, $H(x)$ is the Higgs and $\psi_A(x)$ are two gluinos, $A$ 
is the extended supersymmetry index, $A=1,2$. All fields 
are supposed to be matrices in the adjoint representation of the Lie algebra of the gauge group. 
The trace $\Tr\,\{\cdot\}$ is taken over the adjoint representation and $h^\vee$ 
is the dual Coxeter number of the gauge group. $\Theta$ is the instanton angle, 
the first term is responsible for the strong CP violation in the Standard Model. 
Note that the first term is the topological invariant of the principle bundle
 whose connection is $A_\mu$, it does not affect to the equations of motion and 
 its contribution to the action is $\Theta k$, where $k \in {\mathbb Z}$ is the \emph{instanton number} 
 or, mathematically speaking, the second Chern class.

The action \eqref{MicroAction} can be written in a more compact way using the ${\mathcal N}=2$ 
extended superfield formalism. Let us introduce the ${\mathcal N}=2$ superspace 
whose coordinates are $x \in {\mathbb R}^4$ and $\theta_A^\alpha \in \Pi {\mathbb C}^4$, where $\alpha = 1,2$ 
is the Weyl spinor index and $\Pi {\mathbb C}^4$ is the 4-dimensional complex space 
with fermionic statistics. We introduce the ${\mathcal N}=2$ chiral supermultiplet as follows:
\begin{gather}
\label{N=2Chiral}
\Psi(x,\theta) = H(x) + \sqrt{2} \theta_A\psi^A(x) + \frac{1}{\sqrt{2}} \theta_A \sigma^{\mu\nu}
 \theta^A F_{\mu\nu}(x) + \cdots.
\end{gather}
Then the action \eqref{MicroAction} can be written as
\begin{gather}
\label{MicroActionSuperField}
S_{\rm micro}(X) = \frac{1}{4\pi} \Im{\mathfrak m} \left\{ \int {\rm d}^4 x \, {\rm d}^4 \theta 
\frac{\tau}{2h^\vee} \Tr \Psi^2(x,\theta) \right\},
\end{gather}
where 
\begin{gather}
\label{ComplexCoupling}
\tau = \frac{4\pi i}{g^2} + \frac{\Theta}{2\pi}
\end{gather}
is the \emph{complex coupling constant}.

\subsection{Low energy physics}

At low energies thanks to the Higgs potential $- \frac{1}{2} {[H,H^\dag]}^2$ 
the Higgs field can have non-zero vacuum expectations values (vevs). 
Let us consider the situation when the supersymmetry remains unbroken at low energies. 
It follows that the ground state energy is zero. Therefore we arrive to the condition $[H,H^\dag] = 0$. 
It follows that $\langle  H \rangle$ is the diagonal matrix. Mathematically speaking the vevs 
of Higgs belong to the Cartan subalgebra of the gauge group Lie algebra. If the rank
 of the algebra is $r$ then $\langle H_l \rangle_a = -2\sqrt{2} a_l$, $l=1,\dots,r$ and we denote by 
 $\langle \cdot\rangle_a$ the vacuum expectation over the configuration satisfying such a condition.

If all $a_l$ are different we have the \emph{Coulomb} 
branch of the theory. In this situation we have maximal breaking of the gauge group $G \mapsto {[{\rm U}(1)]}^r / W_G$, 
where $W_G$ is the Weyl group of the gauge group Lie algebra which is responsible for the $a_l$'s permutations.

The terms $\nabla^\mu H^\dag \nabla_\mu H$ and $ - \frac{i}{\sqrt{2}} \psi_A [H^\dag, \psi^A] 
+ \frac{i}{\sqrt{2}} \bar{\psi}^A [H,\bar{\psi}_A]$ 
are responsible for the mass appearance in the gluon and gluinos field. The mass is of order of $a_l - a_m$.
 Therefore in the Coulomb branch  the only massless states belong to the Cartan subalgebra. 

Now let us consider the Wilsonian low-energy effective action defined as
\begin{gather}
\label{Wilsonian}
{\rm e}^{\frac{i}{\hbar} S_{\rm eff}(\tilde{X},\Lambda)} = \int_{|k| > \Lambda} {\mathcal D} X 
\,{\rm e}^{\frac{i}{\hbar} S_{\rm micro}(X)}.
\end{gather}
If $|a_l - a_m| \gg \Lambda$ when $l\neq m$ the $\tilde{X}$ consists of the 
fields contained in the Cartan subalgebra part of $\Psi(x,\theta)$, otherwise in $\Psi_l(x,\theta)$.

The unbroken supersymmetry condition implies the strong restriction on the 
form of the effective action. Namely one can show that \cite{Prepotential}
\begin{gather}
\label{EffectiveAction}
S_{\rm eff}(\Psi,\Lambda) = \frac{1}{4\pi} \Im{\mathfrak m}
\left\{ \frac{1}{2\pi i}\int 
{\rm d}^4 x {\rm d}^4 \theta {\mathcal F}(\Psi,\Lambda) \right\} 
+ \int {\rm d}^4 x{\rm d}^4 \theta {\rm d}^4 \bar{\theta}{\mathcal H}(\Psi,\bar{\Psi},\Lambda) + \cdots,
\end{gather}
where ${\mathcal F}(\Psi,\Lambda)$ is an analytical function of $r+1$ 
variables known as \emph{prepotential}, ${\mathcal H}(\Psi,\bar{\Psi},\Lambda)$ 
is a~real function. The leading term contains up to 2~derivatives and 4~fermions, 
the second contains 4~derivatives and 8~fermions and so on. Note that 
the leading term is the direct generalization of~\eqref{MicroActionSuperField}.

\subsection[Seiberg-Witten solution]{Seiberg--Witten solution}

The leading term of the effective action \eqref{EffectiveAction} 
is the main object of our investigation. It is comp\-letely defined by the 
prepotential ${\mathcal F}(\Psi,\Lambda)$. The prepotential can be split into three parts:
\[
{\mathcal F}(\Psi,\Lambda) = {\mathcal F}^{\rm class}(\Psi) + {\mathcal F}^{\rm pert}(\Psi,\Lambda) 
+ {\mathcal F}^{\rm inst}(\Psi,\Lambda).
\]
The first term is the classical prepotential, which is defined 
by \eqref{MicroActionSuperField} and equals ${\mathcal F}^{\rm class}(\Psi) = \pi i \tau_0 \sum\limits_{l=1}^r \Psi_l^2$.
 The perturbative contributions are purely of 1-loop nature, and therefore 
 can be computed by 1-loop Feynman diagrams. The result is
\begin{gather}
{\mathcal F}^{\rm pert}(\Psi,\Lambda) = - \sum_{\alpha\in \Delta^+} {\boldsymbol k}_\Lambda
\big(\langle \Psi,\alpha\rangle\big) + \frac{1}{2} \sum_{\varrho} \sum_{\lambda \in {\boldsymbol w}_\varrho}
{\boldsymbol k}_\Lambda\big(\langle\Psi,\lambda\rangle + m_\varrho\big) \nonumber\\
\phantom{{\mathcal F}^{\rm pert}(\Psi,\Lambda)}{}=  - \sum_{\alpha\in \Delta^+} {\boldsymbol k}
\big(\langle \Psi,\alpha\rangle\big) 
+ \frac{1}{2} \sum_{\varrho} \sum_{\lambda \in {\boldsymbol w}_\varrho}{\boldsymbol k}\big(\langle\Psi,\lambda\rangle 
+ m_\varrho\big) + \frac{\beta}{2} \ln \Lambda \sum_{l=1}^r \Psi_l^2,\label{PrepPert}
\end{gather}
where 
\begin{gather*}
{\boldsymbol k}_\Lambda(x) = \frac{1}{2}  x^2 \left(\ln \left| \frac{x}{\Lambda}\right| 
- \frac{3}{2}\right), \qquad {\boldsymbol k}(x) = {\boldsymbol k}_1(x) = \frac{1}{2} x^2 
\left( \ln |x| - \frac{3}{2}\right),
\end{gather*}
and $\beta$ in the last line is the leading (and the only) coefficient of the $\beta$-function 
expansion. $\Delta^+$ is the set of all positive roots of the gauge group Lie algebra and ${\boldsymbol w}_\varrho$ 
is the weight system of the matter multiplet representation $\varrho$ of the gauge group, $m_\varrho$ 
being its mass. Combining \eqref{PrepPert} with the classical prepotential 
we get the following RG-flow for the complex coupling constant:
\begin{gather}
\label{RG-flow}
\tau(\Lambda) = \tau_0 + \frac{\beta}{2\pi i} \ln \Lambda.
\end{gather}
Note that $\beta$ in these formulae is always integer.

The classical theory has ${\rm U}(2) = {\rm SU}(2)_I \times {\rm U}(1)_{\mathcal R}$ 
internal symmetry. In the quantum theory due to 
the ABJ anomaly the last factor becomes broken down to ${\mathbb Z}_\beta \equiv {\mathbb Z} / \beta{\mathbb Z}$. 
According to this (thinking of $\Lambda$ as of a vacuum expectation value of 
a supplementary field with the same symmetry~\cite{Prepotential}) we get
\begin{gather*}
{\mathcal F}^{\rm inst}(\Psi,\Lambda) = \sum_{k=1}^\infty {\mathcal F}_k(\Psi) \Lambda^{k\beta}.
\end{gather*}
Now using \eqref{RG-flow} and \eqref{ComplexCoupling} we conclude that 
\begin{gather*}
\Lambda^{k\beta} \sim {\rm e}^{2\pi i k \tau} = {\rm e}^{-\frac{8\pi^2 k}{g^2} + ik \Theta} =
{\rm e}^{-S_{\rm micro}(X_k)},
\end{gather*}
where $X_k$ is the solution of the classical equations of motion for the $k$-instanton sector. 
Therefore the $\Lambda^\beta$-expansion of  ${\mathcal F}^{\rm inst}(\Psi,\Lambda)$ 
can be identified with the instanton expansion, and each term ${\mathcal F}_k(\Psi)$ comes from the $k$-instanton vacuum.

\begin{figure}
\centerline{\includegraphics[width=8cm]{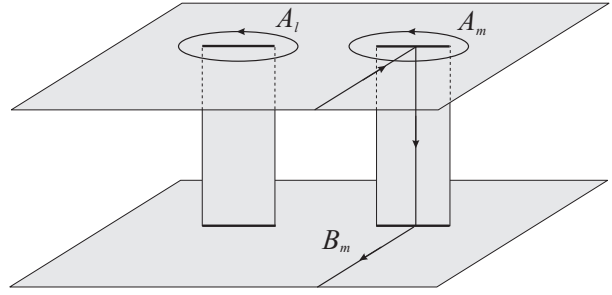}}
\caption{$A$- and $B$-cycles.}\label{Cycles}
\end{figure}

Seiberg and Witten in  \cite{SeibergWitten,SeibergWittenII} 
have proposed a very elegant, but rather indirect way to determine ${\mathcal F}^{\rm pert}(\Psi,\Lambda) 
+ {\mathcal F}^{\rm inst}(\Psi,\Lambda)$. 
The following supplementary objects are needed: an algebraic curve, which is defined as a 
zero locus of a polynomial ${\mathcal C}(z,y,q)$, $q = {\rm e}^{2\pi i \tau}$ 
being the instanton counting parameter, and a meromorphic differential $\lambda(z)$ 
defined in such a way that its derivatives with respect to the algebraic curve
 moduli be holomorphic differentials. Then the prepotential is defined by the following relations
\begin{gather}
\label{SeibergWitten}
a_l = \oint_{A_l} \lambda,  \qquad a_D^l = \frac{\partial {\mathcal F}}{\partial a_l} = 2\pi i\oint_{B_l} \lambda, 
\end{gather}
where $A_l$ and $B_l$ are basic cycles of the algebraic curve, which can be
 represented as a Riemann surface with cuts (see Fig.~\ref{Cycles}). 
 The intersection number is $A_l \# B_m = \delta_{l,m}$.

Seiberg--Witten theory allows to reduce the non-perturbative 
computation problem to a problem of Riemann geometry. However 
to use the whole power of this approach we need to know the exact form 
of the algebraic curve, that is, the polynomial ${\mathcal C}(z,y,q)$. 
Initially for the ${\rm SU}(2)$ case it was done using the first principles. 
However in more general situation it becomes too difficult. Some 
other methods was invented to get access to curves. Let us mention the relation 
with the integrable models~\cite{SYMandIntegr}, the geometrical engineering~\cite{GeomEng,GeomEng2} 
and the Type IIA/M-theory engineering~\cite{PrepFromM,GroupProductFromBranes}. 
In the last approach the Seiberg--Witten curve which was initially introduced 
as an auxiliary object becomes a subspace of the 11-dimensional target space.

\section{Instanton counting}

In this section we describe the Nekrasov approach to ${\mathcal N}=2$ super Yang--Mills theory. 
First we explain how the closed relation for the prepotential can be obtained 
using some peculiar properties of the microscopic action~\eqref{MicroAction}. 
When we show how the Seiberg--Witten curves can be extracted form the exact formulae.

\subsection{Localization and deformation}

In \cite{SWfromInst} a new powerful technique 
to compute the low-energy effective action was proposed. 
The~idea is to perform the \emph{direct} evaluation of 
the functional integral, which defines the partition function 
of the theory using the \emph{localization} approach. As it 
was noticed for the first time in~\cite{TQFT,IntroToCohFT} 
the ${\mathcal N}=2$ super Yang--Mills action is an example of 
the \emph{Cohomological Field Theories}. Namely, one can introduce linear combinations of supercharges 
\begin{gather}
\label{TwistedSupercharges}
\bar{\mathcal Q} = \epsilon^{A\dot{\alpha}} \bar{\mathcal Q}_{A,\dot{\alpha}}, \qquad
 {\mathcal Q}_\mu =  \bar{\sigma}_\mu^{A\alpha} {\mathcal Q}_{A,\alpha}, \qquad
  \bar{\mathcal Q}_{\mu\nu} = \bar{\sigma}_{\mu\nu}{}^{A\dot{\alpha}} \bar{\mathcal Q}_{A,\dot{\alpha}},
\end{gather} 
which, after the topological twist become scalar, vector and anti-selfdual 
two-form with respect to the Lorentz group. The scalar fermionic 
operator can be identified with the BRST operator~\cite{BaulieuSinger} 
for the appropriate gauge fixing procedure for the topological action
\begin{gather}
\label{TopologicalAction}
S_{\rm top} = \frac{\Theta}{32 \pi^2 h^\vee} \int {\rm d}^4 x\, \Tr\left\{F_{\mu\nu} \star F^{\mu\nu} \right\}, 
\end{gather}
where $h^\vee$ is the dual Coxeter number, 
the trace is taken over the adjoint representation and $\star F_{\mu\nu} = \frac{1}{2} \epsilon_{\mu\nu\rho\sigma} 
F^{\rho\sigma}$. 

Another amazing property of the ${\mathcal N}=2$ 
super Yang--Mills action is the existence of a Lorentz 
deformation which preserves one of four supercharges. 
Technically this deformation can be explained as follows. 
The ${\mathcal N}=2$ supersymmetric Yang--Mills action can be 
obtained as the compactification of ${\mathcal N}=1$, $d=6$ super Yang--Mills action. 
If we compactify $x^4 \equiv x^4 + 2\pi R_4$ and 
$x^5 \equiv x^5 + 2\pi R_5$ in the flat 
space ${\rm d} s_6^2 = g_{\mu\nu} {\rm d} x^\mu \, {\rm d} x^\nu - {\rm d} x^a \,
{\rm d} x^a$, $a = 4,5$, we get the undeformed theory. 
However if we use the following spacetime dependent metric~\cite{SmallInst} 
\begin{gather*}
{\rm d} s^2_6 = g_{\mu\nu} \big( {\rm d} x^\mu + V_a^\mu(x) {\rm d} x^a \big)
\big({\rm d} x^\nu + V_b^\nu(x) {\rm d} x^b \big) - {\rm d} x^c \,{\rm d} x^c,
\end{gather*}
where $V_a^\mu(x) = \Omega_{a,\nu}^\mu x^\nu$ and $\Omega^\mu_{a,\nu}$ 
are matrices of Lorentz rotations, we get the deformed version of the theory. 
In this deformed version the dynamically generated scale becomes effectively superspace dependent, 
due to the RG-equation~\eqref{RG-flow} and the following redefinition of the 
complex coupling constant~\eqref{ComplexCoupling}
\begin{gather}
\label{SuperTau}
\tau(x,\theta) = \tau - \frac{1}{\sqrt{2}} \left({(\bar{\Omega}_{\mu\nu})}^+ 
\theta^\mu\theta^\nu - \frac{1}{2\sqrt{2}} \bar{\Omega}_{\mu\nu}\Omega^{\mu}{}_\rho x^\rho x^\nu \right),
\end{gather}
where $\theta_\mu = \bar{\sigma}_\mu^{A\alpha} \theta_{A,\alpha}$ 
is twisted at the same way as \eqref{TwistedSupercharges} supercoordinates and 
\begin{gather*}
\Omega_{\mu\nu} = \frac{1}{\sqrt{2}}\big(\Omega_{4,\mu\nu} + i \Omega_{5,\mu\nu} \big), \qquad
 {(\Omega_{\mu\nu})}^+ = \frac{1}{2}\big(\Omega_{\mu\nu} + \star \Omega_{\mu\nu} \big).
\end{gather*}

The only survived supercharge is the supercharge which 
annihilates the superspace dependent complex coupling constant~\eqref{SuperTau}. It is
\begin{gather*}
\bar{\mathcal Q}_\Omega = \bar{\mathcal Q} + \frac{1}{2\sqrt{2}} \Omega^\mu{}_\nu x^\nu {\mathcal Q}_\mu.
\end{gather*}

It follows that the functional integral which represents the partition function of the theory can be computed as
\begin{gather}
\langle 1 \rangle_a = \int {\mathcal D} X {\rm e}^{-S_{\rm micro}(X)} = \int_{|k| < 
\Lambda} {\mathcal D} \tilde{X} {\rm e}^{-S_{\rm eff}(\tilde{X})} \nonumber\\
\phantom{\langle 1 \rangle_a}{} = \exp \left\{ \frac{1}{4\pi} \Im{\mathfrak m} \frac{1}{2\pi i} \int {\rm d}^4 x \,
{\rm d}^4 \theta {\mathcal F}(-2\sqrt{2}a,\Lambda(x,\theta))\right\} 
= \exp \frac{1}{\varepsilon_1\varepsilon_2} {\mathcal F}(a,\Lambda,\varepsilon_1,\varepsilon_2),\label{Nikita'sFormula}
\end{gather}
where in the last line we have localized the integral over the superspace 
in the origin using the Duistermaat--Heckman formula~\cite{Duistermaat-Heckman} 
(which is also based on the idea of localization), and we have used the fact that 
the prepotential is an homogeneous function of degree~2, also we have renormalized 
the ultraviolet cut-off~\cite{SWfromInst}.

The deformation parameters $\varepsilon_1$ and $\varepsilon_2$ are contained in the Lorentz rotation matrix as follows
\begin{gather*}
\Omega_{\mu\nu} = \frac{1}{\sqrt{2}}\left(
\begin{array}{cccc}
0 & 0 & 0 & \varepsilon_1 \\
0 & 0 & \varepsilon_2 & 0 \\
0 & -\varepsilon_2 & 0 & 0 \\
-\varepsilon_1 & 0 & 0 & 0 
\end{array}
\right).
\end{gather*}

Another way to compute the partition function is to note that 
the microscopic action of the deformed theory is, up to the 
topological term~\eqref{TopologicalAction}, $\bar{\mathcal Q}_{\Omega}$-exact. 
It allows us to localize the whole Feynman 
integral on the instanton moduli space~\cite{TQFT,SWfromInst}. 
The typical expression is given by a~sum over the different instanton  sectors:
\begin{gather}
\label{localization}
\langle 1 \rangle_a = Z^{\rm pert}(a,\Lambda,\varepsilon_1,\varepsilon_2) 
\left( 1 + \sum_{k=1}^\infty q^k Z_k(a,\varepsilon_1,\varepsilon_2) \right),
\end{gather}
where $Z_k(a,\varepsilon_1,\varepsilon_2)$ can be seen as the equivariant 
Euler characteristics of the $k$-instanton moduli space.

The advantage of this method is that by combining~\eqref{Nikita'sFormula} and~\eqref{localization} we get 
the \emph{direct access} to the prepotential, which can be represented by a 
contour integral of a rational function of some auxiliary variables, and can 
be (in principle) computed by residues. For example, the equivariant Euler 
characteristics of the $k$-instanton sector for the pure ${\rm SU}(N)$ 
theory is given by the following expression ($\varepsilon_+ = \frac{\varepsilon_1 + \varepsilon_2}{2}$):
\begin{gather}
Z_k(a,\varepsilon_1,\varepsilon_2) = \frac{1}{k!} 
\frac{{(\varepsilon_1+\varepsilon_2)}^k}{\varepsilon_1^k\varepsilon_2^k} 
\oint \prod_{i=1}^k \frac{{\rm d} \phi_i}{2\pi i} \prod_{i\neq j}
\frac{(\phi_i - \phi_j)(\phi_i - \phi_j - \varepsilon_1 - \varepsilon_2)}
{(\phi_i - \phi_j - \varepsilon_1)(\phi_i - \phi_j - \varepsilon_2)}\nonumber\\
\phantom{Z_k(a,\varepsilon_1,\varepsilon_2) =}{}
\times \prod_{i=1}^k \prod_{l=1}^N \frac{1}{(\phi_i - a_l - \varepsilon_+)(\phi_i - a_l + \varepsilon_+)}.\label{Zk}
\end{gather}
In \cite{SWfromInst} it was shown how to compute this integral while taking into account the combinatorics of residues.

\subsection{Thermodynamical limit}

Even though we have got an exact expression for the prepotential, 
that is, for the low-energy effective action, this expression is a 
series over the ultraviolet cut-off $\Lambda$. However, to study such 
non-perturbative effects as confinement and the monopole and dyon 
condensation we should be able to make an analytical continuation beyond 
the convergence radius. A method which can help us is the Seiberg--Witten theory. 
Recall that the prepotential which is defined by an algebraic curve can be 
defined for any $\Lambda$. Therefore should we have a curve, the continuation can be easily obtained.

So now we are faced to the ``inverse problem'': 
we know the series on $\Lambda$ around zero and we wish to reconstruct the exact Seiberg--Witten curve. 
This problem was solved in~\cite{SWandRP}. The idea is that in 
fact we are interested in the non-deformed theory. Therefore we can 
think of $\varepsilon_1$ and $\varepsilon_2$ as of small parameters. 
One can show that in the limit $\varepsilon_1\varepsilon_2\to 0$ the 
whole sum~\eqref{localization} is dominated by a single term with 
$k\sim \frac{1}{\varepsilon_1\varepsilon_2} \to \infty$. This effect can
be illustrated by the following example: consider the series for the exponent ${\rm e}^x$. 
When $x\to \infty$ the whole sum is dominated by a single term with $k \sim x$. Indeed, we have
\begin{gather*}
{\rm e}^x = \sum_{k=0}^\infty \frac{x^k}{k!} \sim \frac{x^x}{x!},
\end{gather*}
which is another way to claim the Stirling's formula.

Another observation is that when $k\to\infty$ the $k$-tuple 
integration can be replaced by the functional integration over the instanton density
\begin{gather*}
\rho(x) = \varepsilon_1\varepsilon_2 \sum_{i=1}^k\delta (x-\phi_i).
\end{gather*}
Formulae get simple if instead of the instanton density we introduce the \emph{profile function}
\begin{gather*}
f(x) = \sum_{l=1}^N |x-a_l| - 2 \rho(x).
\end{gather*}
This function possess the following features (which can be proved quite straightforwardly)
\begin{gather}
\frac{1}{2}\int_{\mathbb R} {\rm d} x \, f''(x) = N, \qquad  \frac{1}{2}
\int_{\mathbb R} {\rm d} x\,  f''(x) x = \sum_{l=1}^N a_l = 0, \nonumber\\
\frac{1}{2}  \int_{\mathbb R} {\rm d} x \, f''(x) x^2 = \sum_{l=1}^N a_l^2 - 2 \varepsilon_1\varepsilon_2 k. \label{f-prop}
\end{gather}
The last equation fixes the relation between $k$ and $\varepsilon_1\varepsilon_2$.

Then the partition function can be represented in this limit as follows:
\begin{gather*}
\langle 1 \rangle_a \sim \int {\mathcal D} f \exp\left\{-\frac{1}{\varepsilon_1\varepsilon_2}
 \left( H[f] + \frac{\pi i \tau(\Lambda)}{2} \int {\rm d} x \, f''(x) x^2  + O(\varepsilon_1,\varepsilon_2)\right) 
 \right\}.
\end{gather*}
The Hamiltonian $H[f]$ can be reconstructed with the help of the exact formula 
for the prepotential. Consider, for example, the ${\rm SU}(N)$ model without matter hypermultiplets. 
The exact expression is given by~\eqref{Zk}. We have
\begin{gather*}
\prod_{i\neq j} \frac{(\phi_i - \phi_j)(\phi_i - \phi_j - \varepsilon_1 - \varepsilon_2)}
{(\phi_i - \phi_j - \varepsilon_1)(\phi_i - \phi_j - \varepsilon_2)} 
\sim \exp \left(-\varepsilon_1\varepsilon_2 \sum_{\i \neq j} \frac{1}{{(\phi_i - \phi_j)}^2} \right) \\ 
\qquad {} = \exp\left( - \frac{1}{\varepsilon_1\varepsilon_2} \int {\rm d} x \, {\rm d} y  \,
\frac{\rho(x)\rho(y)}{{(x-y)}^2}\right) = \exp \left(\frac{1}{\varepsilon_1\varepsilon_2} \int {\rm d} x\,
 {\rm d} y \, \rho''(x)\rho''(y) {\boldsymbol k}(x-y) \right), 
\\
\prod_{i=1}^k \prod_{l=1}^N \frac{1}{(\phi_i - a_l - \varepsilon_+)
(\phi_i - a_l + \varepsilon_+)} \sim \exp \left(-2\sum_{i=1}^k \sum_{l=1}^N \ln (\phi_i - a_l)\right) \\ 
\qquad {}= \exp \left( -\frac{2}{\varepsilon_1\varepsilon_2} \sum_{l=1}^N \int {\rm d} x \, \rho(x) 
\ln (x-a_l)\right) = \exp\left( -\frac{2}{\varepsilon_1\varepsilon_2} \sum_{l=1}^N \int {\rm d} x\,
 \rho''(x) {\boldsymbol k}(x-a_l)\right).
\end{gather*}
Using \eqref{PrepPert}, \eqref{RG-flow} and \eqref{f-prop} the perturbative 
contribution to the partition function can be rewritten as follows 
\begin{gather*}
Z^{\rm pert}(a,m,\Lambda;\varepsilon) q^k = Z^{\rm pert}(a,m,\Lambda;\varepsilon) \Lambda^{\beta k} {\rm e}^{2\pi i k 
\tau_0} \\
\qquad {}= \exp \frac{1}{\varepsilon_1\varepsilon_2} \left( {\mathcal F}^{\rm class}(a,m) 
+ {\mathcal F}^{\rm pert}(a,m,1)
 - \frac{\pi i}{2}\tau(\Lambda) \int {\rm d} x \, f''(x) x^2 + O(\varepsilon_1,\varepsilon_2) \right).
\end{gather*}
Having combined these three formulae we finally get 
\begin{gather*}
q^k Z^{\rm pert}Z_k \sim \int {\mathcal D} f \exp \left\{ - \frac{1}{\varepsilon_1\varepsilon_2}
 \left(- \frac{1}{4} \int {\rm d} x \, {\rm d} y\,  f''(x) f''(y) {\boldsymbol k}(x-y) \right.\right.\\
\left.\left. \phantom{q^k Z^{\rm pert}Z_k \sim}{}
 + \frac{\pi i \tau(\Lambda)}{2} \int {\rm d} x\, f''(x) x^2 \right)\right\}.
\end{gather*}

The steps can be performed in the general case. We have put corresponding contributions to 
the Hamiltonians $H[f]$ into the Table~\ref{Hams}, $m$ being the mass of the matter hypermultiplet. 

\begin{table}[t]
\begin{center}
\caption{Hamiltonians.}\label{Hams}
\vspace{2mm}
\small 
\begin{tabular}{||c|c||c||}
\hhline{|t:=:=:t:=:t|}
&&\\[-3.0mm]
\textbf{Group} & \textbf{Multiplet} & \textbf{Contribution to  ${\boldsymbol H[f]}$}\\[0.5mm]
\hhline{|:=:=::=:|} 
&&\\[-3.0mm] 
& Adjoint, gauge & $\displaystyle  -\frac{1}{4}\int {\rm d} x{\rm d} y f''(x)f''(y){\boldsymbol k}(x-y)$\\[0.5mm]
\hhline{||~|-||-||}
&&\\[-3.0mm] 
& Fundamental & $\displaystyle   \frac{1}{2} \int {\rm d} x f''(x) {\boldsymbol k}(x+m)$\\[0.5mm]
\hhline{||~|-||-||}
&&\\[-3.0mm] 
${\rm SU}(N)$ & Symmetric & $\displaystyle  \frac{1}{8}\int {\rm d} x{\rm d} y f''(x)f''(y) {\boldsymbol k}(x+y + m) +  \int {\rm d} x f''(x) {\boldsymbol k}(x + m/2)$ \\[0.5mm]
\hhline{||~|-||-||}
&&\\[-3.0mm] 
& Antisymmetric & $\displaystyle  \frac{1}{8} \int {\rm d} x{\rm d} y f''(x)f''(y){\boldsymbol k}(x + y + m) -  \int {\rm d} x f''(x) {\boldsymbol k}(x + m/2)$\\[0.5mm]
\hhline{||~|-||-||}
&&\\[-3.0mm] 
& Adjoint, matter & $\displaystyle  \frac{1}{4} \int {\rm d} x{\rm d} y f''(x)f''(y){\boldsymbol k}(x - y + m)$\\[0.5mm]
\hhline{|:=:=::=:|}
&&\\[-3.0mm] 
& Adjoint, gauge & $\displaystyle -\frac{1}{8} \int {\rm d} x{\rm d} y f''(x)f''(y){\boldsymbol k}(x + y) + \int {\rm d} x f''(x) {\boldsymbol k}(x)$\\[0.5mm]
\hhline{||~|-||-||}
&&\\[-3.0mm] 
${\rm SO}(N)$ & Fundamental & $\displaystyle  \frac{1}{2} \int {\rm d} x f''(x) {\boldsymbol k}(x+m)$\\[0.5mm]
\hhline{||~|-||-||}
&&\\[-3.0mm] 
& Adjoint, matter & $\displaystyle  \frac{1}{8} \int {\rm d} x{\rm d} y f''(x)f''(y){\boldsymbol k}(x + y + m) -  \int {\rm d} x f''(x) {\boldsymbol k}(x + m/2)$ \\[0.5mm]
\hhline{|:=:=::=:|}
&&\\[-3.0mm] 
& Adjoint, gauge & $\displaystyle  -\frac{1}{8} \int {\rm d} x{\rm d} y f''(x)f''(y){\boldsymbol k}(x + y) - \int {\rm d} x f''(x){\boldsymbol k}(x)$\\[0.5mm]
\hhline{||~|-||-||}
&&\\[-3.0mm] 
${\rm Sp}(N)$ & Fundamental & $\displaystyle  \frac{1}{2} \int {\rm d} x f''(x) {\boldsymbol k}(x+m)$\\[0.5mm]
\hhline{||~|-||-||}
&&\\[-3.0mm] 
& Antisymmetric & $\displaystyle \frac{1}{8} \int {\rm d} x{\rm d} y f''(x)f''(y){\boldsymbol k}(x + y + m)  - \int {\rm d} x f''(x) {\boldsymbol k}(x + m/2)$\\[0.5mm]
\hhline{||~|-||-||}
&&\\[-3.0mm] 
& Adjoint, matter & $\displaystyle \frac{1}{8}\int {\rm d} x{\rm d} y f''(x)f''(y) {\boldsymbol k}(x+y + m)  + \int {\rm d} x f''(x){\boldsymbol k}(x+m/2)$\\[0.5mm]
\hhline{|b:=:=:b:=:b|}
\end{tabular}
\end{center}\vspace{-5mm}
\end{table}

After all manipulations we arrive to  the following picture: the main contribution 
to the prepotential is given by the minimizer $f_\star(x)$ 
of the Hamiltonian. The supporter of this minimizer is a union 
of some disjoint intervals $\gamma_l \ni a_l$, and on these intervals the following equation holds:
\begin{gather*}
\frac{1}{\pi i} \frac{\delta H[f]}{\delta f'(t)} = \xi_l + t \tau(\Lambda), \qquad  t \in \gamma_l,
\end{gather*}
where $\tau(\Lambda)$ is given by \eqref{RG-flow}.

This equation can be recast into a \emph{difference equation} for the primitive of the partition function resolvent
\begin{gather*}
F(z) = \frac{1}{4\pi i}\int_{\mathbb R} {\rm d} x \, f''(x) \ln(z-x).
\end{gather*}

The solution of the difference equation allows us to reconstruct the Seiberg--Witten 
curve and the Seiberg--Witten differential as follows. The curve ${\mathcal C}(z,y,q)$ 
and the differential $\lambda(z)$ are given by the dependence $y(z,q)$, where 
\begin{gather*}
y(z,q) = \exp 2\pi i F(z,q), \qquad  \lambda(z) = \frac{1}{2\pi i} z \frac{{\rm d} y(z)}{y(z)} = z {\rm d} F(z,q).
\end{gather*}

The expressions for the Hamiltonians are put into the Table~\ref{Hams}.

\begin{table}
\small 
\caption{Models accepted by the asymptotic freedom condition.}\label{list}

\vspace{-2mm}

\begin{itemize}
\item \underline{${\rm SU}(N)$}:
\begin{itemize}
\itemsep=0pt
\item $N_f$ fundamental multiplets, $N_f \leq 2N$,
\item 1 antisymmetric multiplet and $N_f$ fundamentals, $N_f \leq N + 2$,
\item 1 antisymmetric multiplet and $N_f$ fundamentals, $N_f \leq N-2$,
\item 2 antisymmetric multiplets and $N_f$ fundamentals, $N_f \leq 4$,
\item 1 symmetric multiplet and 1 antisymmetric,
\item 1 adjoint multiplet.
\end{itemize}
\item \underline{${\rm SO}(N)$}:
\begin{itemize}
\itemsep=0pt
\item $N_f$ fundamental multiplets, $N_f \leq N-2$,
\item 1 adjoint multiplet.
\end{itemize}
\item \underline{${\rm Sp}(N)$}:
\begin{itemize}
\itemsep=0pt
\item $N_f$ fundamental multiplets, $N_f \leq N+2$,
\item 1 antisymmetric multiplet and $N_f$ fundamental, $N_f \leq 4$,
\item 1 adjoint multiplet.
\end{itemize}
\end{itemize}
\vspace{-5mm}

\end{table}

\begin{table}[t]
\begin{center}
\small
\caption{Dualities.}\label{dualities}
\vspace{2mm}

\begin{tabular}{||c|c||c|c|c||}
\hhline{|t:=:=:t:=:=:=:t|}
\textbf{Group} & \textbf{Multiplet} & \textbf{Higgs} & \textbf{Fund.} & \textbf{Anti.} \\
\hhline{|:=:=::=:=:=:|}
${\rm SU}(N)$ & Symmetric, $m$ & $\displaystyle  \vec{a}$ &  $m/2$, $m/2$, $m/2$, $m/2$ & $m$ \\
\hhline{|:=:=::=:=:=:|}
& Adjoint, gauge & $\displaystyle  (\diamondsuit,\vec{a},-\vec{a})$ &0, 0, 0, 0 & $-$ \\
\hhline{||~|-||-|-|-||}
${\rm SO}(N)$ & Fundamental, $m$ & $\displaystyle  (\diamondsuit,\vec{a},-\vec{a})$ & $-m$, $+m$  & $-$ \\ 
\hhline{||~|-||-|-|-||}
& Adjoint, $m$ & $\displaystyle  (\diamondsuit,\vec{a},-\vec{a})$ & $-$ & $+m$, $-m$ \\
\hhline{|:=:=::=:=:=:|}
& Adjoint, gauge & $\displaystyle  (0,0,\vec{a},-\vec{a})$ & $-$ &  $-$ \\
\hhline{||~|-||-|-|-||}
& Adjoint, gauge &  &  & \\
${\rm Sp}(N)$& + 2 fund., $m=0$ & $\displaystyle  (\vec{a},-\vec{a})$ & $-$ & $-$ \\
\hhline{||~|-||-|-|-||}
& Fundamental, $m$ & $\displaystyle  (\vec{a},-\vec{a})$ & $+m$, $-m$ & $-$ \\ 
\hhline{||~|-||-|-|-||}
& Antisymmetric, $m$ & $\displaystyle  (\vec{a},-\vec{a})$ & $-$ & $+m$, $-m$ \\
\hhline{||~|-||-|-|-||}
& Adjoint, $m$  &$\displaystyle  (\vec{a},-\vec{a})$  & $+m/2$, $+m/2$, $-m/2$, $-m/2$ & $+m$, $-m$  \\
\hhline{|b:=:=:b:=:=:=:b|}
\end{tabular}
\vspace{-6mm}
\end{center}

\end{table}

\section{Results}

Now let us briefly discuss obtained results \cite{SWfromInst,SWandRP,ABCD,SPinSW,MyThesis,CubicCurves}. 
First of all let us mention that for all cases allowed by the asymptotic freedom (Table~\ref{list}) 
the integral expressions similar to~\eqref{Zk} are obtained. In some cases~\cite{SWandRP} 
these integrals can be computed by residues. However in the general case the 
combinatorics of residues is too complicated. The discussion about what is happening in 
the case of ${\rm SO}(N)$, ${\rm Sp}(N)$ and antisymmetric representation of ${\rm SU}(N)$ can be found in~\cite{WyllMar}. 

However for all considered cases the difference equations for $F(z)$ are obtained. 
In some cases they are solved explicitly. Moreover an approximative method which allows 
to provide the 1-instanton correction is developed. It is shown that it 
is consistent with the localization results at the 1-instanton level. 

Both the exact solutions and the 1-instanton approximations are checked 
against the known expression for the Seiberg--Witten curve. The 1-instanton corrections was extracted from Seiberg--Witten curves in~\cite{EnnesMasterFunc,MTheoryTested} (see also references therein).

Even in the case when the exact solution of the model is not known, 
we can claim that if the difference equations which define the curve are 
the same (up to some redefinition of parameters) the same is true for their solutions. 
Otherwise there is a number of \emph{dualities}. The list of 
such dualities is given by the Table~\ref{dualities}. As a ``reference point'' 
we have chosen the ${\rm SU}(N)$ models with a number of antisymmetric and fundamental matters. 
For ${\rm SO}(N)$ the notation $\diamondsuit$ 
is 0 when $N$ is odd and is absent when $N$ is even.
Similar duality table was constructed in~\cite{EllipticMod} after examinating the 1-instanton corrections which follow from Seiberg-Witten curves obtained by M-theory ingeneering method.

\section{Further directions}
Now let us announce some questions which remain unsolved through our investigation:
\begin{itemize}
\itemsep=-1pt
\item Find a recurrent procedure to reconstruct the whole curve starting from the functional equations.
\item Find a method to solve them at once.
\item The prepotential is represented as a contour 
integral which can in principle be done by residues. 
How to do it (how to handle the combinatorics of the residues)?
\item Find a procedure to obtain subleading terms in $\varepsilon_1$, 
$\varepsilon_2$ development of $\langle 1 \rangle_a$. They represent 
the interaction of the gauge theory with weak graviphoton field. 
\end{itemize}

\subsection*{Acknowledgements}

This work was partially supported by the EU MRTN-CT-2004-005104 grant ``Forces Universe'' and by by the MIUR contract no. 2003023852.

\LastPageEnding


\begin{thebibliography}{99}
\footnotesize

\bibitem{BaulieuSinger}
Baulieu L., Singer I.M., Topological Yang--Mills symmetry,
{\it Nucl. Phys.~B, Proc. Suppl.}, 1988, V.5, 12--19.

\bibitem{Duistermaat-Heckman}
 Duistermaat J.J.,  Heckman G.J., On the variation in the cohomology in the
  symplectic form of the reduced phase space, {\it Invent. Math.}, 1982,
  V.69, 259--269.

\bibitem{MTheoryTested}
I.~Ennes, C.~Lozano, S.~Naculich, H.~Rhedin, and H.~Schnitzer, {M}-theory
  tested by {$\mathcal{N}=2$} {S}eiberg-{W}itten theory,
  hep-th/0006141.

\bibitem{EnnesMasterFunc}
I.~Ennes, S.~Naculich, H.~Rhedin, and H.~Schnitzer, Tests of {M}-theory from
  {$\mathcal{N}=2$} {S}eiberg-{W}itten theory,
  hep-th/9911022.

\bibitem{EllipticMod}
Isabel~P. Ennes, Carlos Lozano, Stephen~G. Naculich, and Howard~J. Schnitzer,
  Elliptic models and {M}-theory, \emph{Nucl. Phys.},~2000, V.~B576, 313--346,
  hep-th/9912133.

\bibitem{GroupProductFromBranes}
Erlich J., Naqvi A., Randall L., The Coulomb branch of
  $\mathcal{N} = 2$ supersymmetric product group theories from branes,
  {\it Phys. Rev.~D}, 1998, V.58, 046002, 10~pages;
  hep-th/9801108.

\bibitem{GeomEng2}
Katz S., Mayr P., Vafa C., Mirror symmetry and exact solution of $4D$ $\mathcal{N} = 2$
  gauge theories. I, {\it Adv. Theor. Math. Phys.}, 1998, V.1, 53--114;
  hep-th/9706110.

\bibitem{GeomEng}
Katz S., Klemm A., Vafa C., Geometric engineering of quantum
  field theories, {\it Nucl. Phys.~B}, 1997, V.497, 173--195; hep-th/9609239.

\bibitem{SmallInst}
Losev A., Marshakov A., Nekrasov N., Small instantons, little strings and
  free fermions, hep-th/0302191.

\bibitem{WyllMar}
Mari{\~n}o M., Wyllard N., A note on instanton counting for
  $\mathcal{N} = 2$ gauge theories with classical gauge groups,
  {\it JHEP}, 2004, V.0405, paper 021, 23 pages; hep-th/0404125.

\bibitem{SWfromInst}
Nekrasov N., Seiberg--Witten prepotential from instanton counting,
{\it Adv. Theor. Math. Phys.}, 2004, V.7, 831--864;
hep-th/0206161.

\bibitem{SWandRP}
Nekrasov N., Okounkov A., Seiberg--Witten theory and random partitions,
hep-th/0306238.

\bibitem{ABCD}
Nekrasov N., Shadchin S., ABCD of instantons,
{\it Comm. Math. Phys.}, 2004, V.253, 359--391; hep-th/0404225.

\bibitem{Prepotential}
Seiberg N., Supersymmetry and nonperturbative beta functions, {\it Phys.  Lett.~B}, 1988, V.206, 75--87.

\bibitem{SeibergWitten}
Seiberg N., Witten E., Electric-magnetic duality, monopole condensation, and
  confinement in $\mathcal{N} = 2$ supersymmetric Yang--Mills theory,
{\it Nucl. Phys.~B}, 1994, V.426, 19--52; hep-th/9407087.

\bibitem{SeibergWittenII}
Seiberg N., Witten E., Monopoles, duality and chiral symmetry breaking in $\mathcal{N}=2$
  supersymmetric QCD, {\it Nucl. Phys.~B}, 1994, V.431, 484--550; hep-th/9408099.

\bibitem{CubicCurves}
Shadchin S., Cubic curves from instanton counting,
hep-th/0511132.

\bibitem{MyThesis}
Shadchin S., On certain aspects of string theory/gauge theory correspondence, PhD Thesis, 
Universit\'e Paris-Sud, Orsay, France, 2005;
hep-th/0502180. 

\bibitem{SPinSW}
Shadchin S., Saddle point equations in Seiberg--Witten theory,
{\it JHEP}, 2004, V.0410, paper 033, 38~pages; hep-th/0408066.

\bibitem{IntroToCohFT}
Witten E., Introduction to topological quantum field theories, Lectures at the
  Workshop on Topological Methods in Physics, ICTP, Trieste, Italy (June 1990).

\bibitem{TQFT}
Witten E., Topological quantum field theory, {\it Comm. Math. Phys.}, 1988,
  V.117, 353--386.

\bibitem{PrepFromM}
Witten E., Solutions of four-dimensional field theories via {M}-theory,
{\it Nucl. Phys.~B}, 1997, V.500, 3--42; hep-th/9703166.

\bibitem{SYMandIntegr}
Witten E., Donagi R., Supersymmetric Yang--Mills systems and integrable
  systems, {\it Nucl. Phys. B}, 1996, V.460, 299--334; hep-th/9510101.

\end{thebibliography}
\end{document}